\documentclass[twocolumn,amsmath,amssymb]{revtex4}
\usepackage{graphicx}
\usepackage{dcolumn}
\usepackage{bm}
\begin{document}

\title{Fractal Butterflies of Chiral Fermions in Bilayer Graphene: Phase Transitions 
and Emergent Properties}
\author{Areg Ghazaryan and Tapash Chakraborty$^\ddag$}
\affiliation{Department of Physics and Astronomy,
University of Manitoba, Winnipeg, Canada R3T 2N2}

\date{\today}
\begin{abstract}
We report on our studies of fractal butterflies in biased bilayer graphene in the
fractional quantum Hall effect (FQHE) regime. We have considered the case when the external 
periodic potential is present in one layer and have illustrated the effect of varying 
both the periodic potential strength and the bias voltage on the FQHE and the butterfly
energy gaps. Interestingly, the butterfly spectra exhibits remarkable phase transitions
between the FQHE gap and the butterfly gap for chiral electrons in bilayer 
graphene, by varying either the periodic potential strength or the bias voltage. We also
find that, in addition to those phase transitions, by varying the bias voltage one can 
essentially control the periodic potential strength experienced by the electrons.
\end{abstract}
\maketitle

Graphene \cite{graphene_book,abergeletal} placed on a hexagonal substrate with a twist is 
expected to display the Hofstadter butterfly pattern \cite{hofstadter,expt_old} 
in the energy spectrum of non-interacting Dirac fermions, when subjected to a perpendicular 
magnetic field. The unique butterfly pattern was indeed discovered in 2013 in monolayer and 
bilayer graphene \cite{dean_13,hunt_13,geim_13}, placed on a hexagonal boron nitride substrate 
with rotational misalignment between graphene and the substrate. That arrangement resulted in 
the Moir{\'e} pattern which actually introduces a large-scale periodicity in the Hamiltonian 
of the system, and the fractal butterfly pattern was the result of splitting of the Moir{\'e} 
minibands (secondary Dirac cones) by the magnetic field that are exhibited in the 
magnetoconductance probe \cite{graphene_butterfly}. Interestingly, a mathematical proof of the 
presence of the {\it fractal} pattern in the butterfly spectrum (the `Ten Martini Problem') is 
also available in the literature \cite{ten_martini}. After that exciting experimental discovery 
of the fractal butterflies, more recent studies (both theoretical \cite{apalkov_14} and experimental 
\cite{geim_14}) have focused on the influence of the electron-electron interaction on the 
butterfly spectrum. Electronic properties of Dirac fermions in monolayer and bilayer graphene 
have been exhaustively studied in recent years \cite{graphene_book,abergeletal,FQHE_chapter}. In 
a strong perpendicular magnetic field, interacting Dirac fermions \cite{interaction} display the 
fractional quantum Hall Effect (FQHE) states \cite{FQHE_book} in monolayer \cite{mono_FQHE} and 
bilayer graphene \cite{FQHE_chapter,bi_FQHE}, that has also been experimentally observed 
\cite{FQHE_expt}. The interaction effects in the fractal butterflies are, of course, more complex 
in the fractional quantum Hall effect regime, where one observes an interplay between the quantum 
Hall effect gap and the Hofstadter gap \cite{areg_butterfly}. Interestingly, in this work we
find that the butterfly spectra exhibit remarkable phase transitions for chiral electrons 
in bilayer graphene, where in addition to the phase transitions between the Hofstadter gap and the 
FQHE gap, one can essentially control the periodic potential strength experienced by the 
electrons by varying the bias voltage.

We consider bilayer graphene with Bernal (AB) stacking in an external periodic potential 
with square symmetry \cite{apalkov_14,review,vidar,ando}. We label the layers of bilayer
graphene by the indices 1,2 and assume that the periodic potential is present only in layer 1. 
Considering that the layers are stacked such that the $B^{}_1$ and the $A^{}_2$ sites are 
vertically aligned, the single-particle Hamiltonian of this system in a magnetic 
field (without the periodic potential) is written  
\cite{graphene_book,abergeletal,FQHE_chapter,bilayer}
\begin{equation}
{\cal H}^{bi}_\xi=\xi\left(\begin{array}{cccc} \frac{U}{2} & v^{}_F\pi^{}_- & 0 & 0 \\ 
v^{}_F\pi^{}_+ & \frac{U}{2} & \xi\gamma^{}_1 & 0 \\
0 & \xi\gamma^{}_1 & -\frac{U}{2} & v^{}_F\pi^{}_- \\ 0 & 0 & v^{}_F\pi^{}_+ & 
-\frac{U}{2} \end{array}\right),
\label{SBHamiltonian}
\end{equation}
where $\pi^{}_\pm=\pi^{}_x\pm i\pi^{}_y$, ${\bm \pi}=\mathbf p +e\mathbf A/c$, $\mathbf p$ 
is the two-dimensional electron momentum, $\mathbf A=(0,Bx,0)$ is the vector potential, 
$v^{}_F\approx 10^6\,\mathrm{m/s}$ is the Fermi velocity in graphene, $U$ is the 
inter-layer bias voltage, $\gamma^{}_1\approx0.4$ eV is the interlayer hopping integral 
and $\xi=1$ for $K$ valley and $\xi=-1$ for $K'$ valley. The corresponding wave function 
is described by a four-component spinor $(\psi^{}_{A^{}_1},\psi^{}_{B^{}_1},\psi^{}_{A^{}_2},
\psi^{}_{B^{}_2})^{T}$ for valley $K$ and $(\psi^{}_{B^{}_2},\psi^{}_{A^{}_2},\psi^{}_{B^{}_1},
\psi^{}_{A^{}_1})^{T}$ for valley $K'$, where $\psi^{}_A$ and $\psi^{}_B$ are wave 
functions of sublattices A and B, respectively. We consider the fully spin-polarized electron 
system and therefore disregard the Zeeman energy. The eigenfunction of the Hamiltonian 
(\ref{SBHamiltonian}) then has the form
\begin{equation}
\label{BaseEig}
\Psi^{}_{n,j}=\left(\begin{array}{c}\xi C^{}_1\varphi^{}_{n-1,j} \\ C^{}_2\varphi^{}_{n,j}\\
C^{}_3\varphi^{}_{n,j}\\ \xi C^{}_4\varphi^{}_{n+1,j}
\end{array}\right),
\end{equation}
where $C^{}_1, C^{}_2, C^{}_3, C^{}_4$ are constants and $\varphi^{}_{n,j}$ is the electron 
wave function in the $n$-th Landau level (LL) with the parabolic dispersion, taking into 
account the periodic boundary conditions (PBC) \cite{FQHE_book,note}. In the wave function 
(\ref{BaseEig}) the LL index $n$ can take the values $-1,0,1,\dots$ and we assume that if 
the LL index of $\varphi^{}_{n,j}$ is negative then it is identically equal to zero. In 
this case, for $n=-1$ the wave function (\ref{BaseEig}) is $\Psi^{}_{-1,j}=(0,0,0,
\varphi^{}_{0,j})$ and there is only one energy level corresponding to this case. For $n=0$, 
$C^{}_1=0$ and there are three energy states in this case. Following the convention for 
the indexing the energy levels introduced in \cite{FQHE_chapter}, we label the states for 
$n=-1$ and $n=0$ as $0^{(\xi)}_i$, where $i=-2,-1,1,2$ is the label of states in ascending 
order of the energy values. In particular, for valley $K$ the state corresponding to $n=-1$ has 
the index $0^{(+)}_{-1}$ and $0^{(-)}_{1}$ for valley $K'$.

\begin{figure}
\includegraphics[width=9cm]{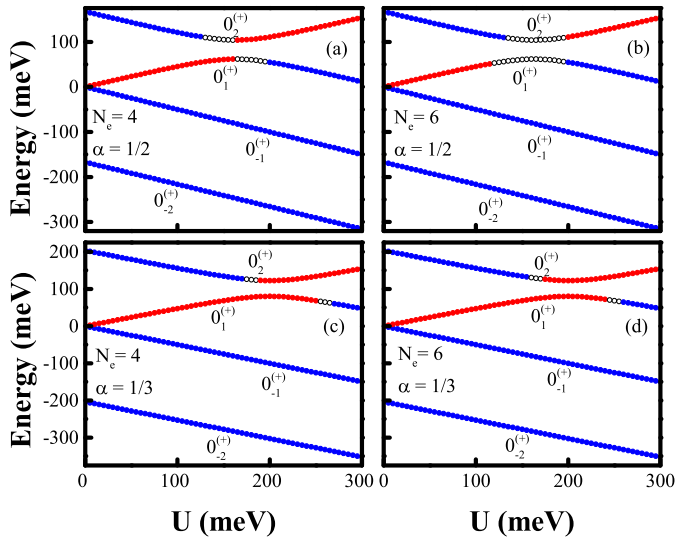}
\caption{\label{fig:LLFQHE} The Landau levels for $n=-1$ and $n=0$ as a function of the bias 
voltage U for two values of $\alpha$, with periodic potential strength $V^{}_0=20$ meV. 
The numbers next to the curves denote the corresponding Landau level as described in the 
text. The regions where the gap corresponds to the FQHE (butterfly) gap are drawn as blue (red) 
dots. (a) and (c) correspond to the system with $N^{}_e=4$ and (b) and (d) to 
$N^{}_e=6$.}\end{figure}

The complete many-body Hamiltonian for this system can be written as 
\begin{equation}
{\cal H}=\sum_i^{N^{}_e}\left[{\cal H}^{bi}_{\xi} + V(x^{}_i,y^{}_i)\right] +
\frac12\sum_{i\neq j}^{N^{}_e}V^{}_{ij}
\label{MBHamiltonian}
\end{equation}
where the second term is the periodic potential, which is nonzero only for the components 
of layer 1 and the last term is the Coulomb interaction. We disregard the valley mixing terms 
(short-range interaction energies) due to the periodic potential and the Coulomb interaction
\cite{areg_butterfly,areg_iqhe}. Both the inter-layer bias voltage and the periodic potential 
break spatial inversion symmetry and therefore the valley degeneracy is lifted in this system 
\cite{Koshino}. For the many-body problem we consider a system of finite number $N^{}_e$ of 
electrons in a toroidal geometry, i.e., the size of the system is $L^{}_x=M^{}_xa^{}_0$ and 
$L^{}_y= M^{}_ya^{}_0$ ($M^{}_x$ and $M^{}_y$ are integers, $a^{}_0$ is the period of the 
external potential) and apply periodic boundary conditions (PBC) in order to eliminate the 
boundary effects. Defining the parameter $\alpha=\phi^{}_0/\phi$ (the inverse of the magnetic 
flux through the unit cell measured in units of the flux quantum), where $\phi=Ba_0^2$ is the 
magnetic flux through the unit cell of the periodic potential and $\phi^{}_0=hc/e$ the 
flux quantum, we have $N^{}_s/(M^{}_xM^{}_y)=1/\alpha,$ where $N^{}_s$ is the number of 
magnetic flux quanta passing through the system or, alternatively, it describes the LL 
degeneracy for each value of the spin and valley index and takes integer values. The filling 
factor is defined as $\nu=p/q=N^{}_e/N^{}_s$, where $p$ and $q$ are again coprime integers. 
In order to solve this problem we first construct the Hamiltonian matrix using the Hamiltonian 
operator (\ref{MBHamiltonian}) and the many-body states $|j^{}_1,j^{}_2,\ldots,j^{}_{N^{}_e}
\rangle$ (besides $j^{}_i$, each single-particle state is characterized by the LL and valley 
indices which are not shown, but are implicitly assumed to be included in the indices $j^{}_i$) 
which are constructed from the single-particle eigenvectors (\ref{BaseEig}). After that we use 
the center of mass (CM) translation algebra \cite{areg_butterfly,areg_iqhe,haldane_85,read,FQHE_book} 
and the eigenstates of CM translations to bring the Hamiltonian matrix into block diagonal 
form, where each block can then be diagonalized using the exact diagonalization procedure 
\cite{FQHE_book}. 

\begin{figure}
\includegraphics[width=8cm]{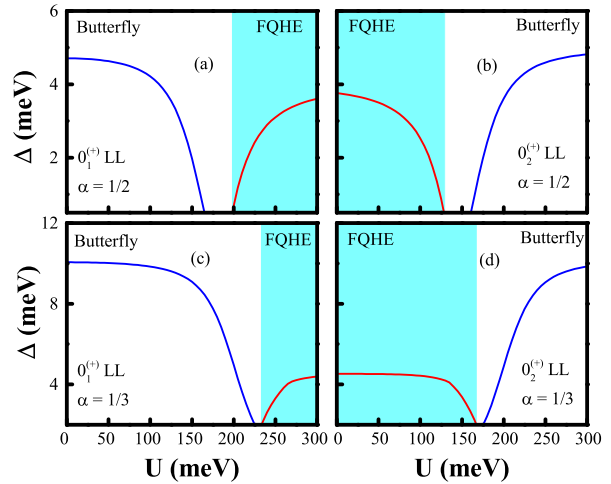}
\caption{\label{fig:GapdepU} The dependence of both the FQHE and the butterfly gaps for 
$N^{}_e=4$ electrons on the bias voltage $U$ and the Landau levels $0^{(+)}_{1}$ and $0^{(+)}_{2}$ 
for two values of $\alpha$, with the periodic potential strength $V^{}_0=20$ meV. The line 
corresponding to the FQHE (butterfly) case is depicted in red (blue). The region of the 
bias voltage $U$ where the gap corresponds to the FQHE gap is marked in cyan. 
}
\end{figure}

In this work we consider the filling factor $\nu=1/3$ for a fully spin polarized system 
and for $\alpha=1/2$ and $\alpha=1/3$. The period of the external potential is taken to be
$a^{}_0=20$ nm, and the inter layer hopping integral to be $\gamma^{}_1=30$ meV 
\cite{bi_FQHE}, which can be achieved by applying an in-plane magnetic field \cite{FQHE_chapter}. 
In Fig.~\ref{fig:LLFQHE} the dependence of the single-particle LL $0^{(+)}_i$ on the
bias-voltage is presented for the $K$ valley. The color of the filled dots indicate the 
regions of the bias voltage where for $N^{}_e=4$ (a,c) and $N^{}_e=6$ (b,d) the gap 
corresponds to the butterfly region or the FQHE region. The periodic potential strength
$V^{}_0$ is taken to be $V^{}_0=20$ meV. In order to understand the phase transitions observed 
in Fig.~\ref{fig:LLFQHE} the wave functions and also the impact of the bias voltage on 
these wave functions should be analyzed for each LL. As mentioned above, for level $0^{(+)}_{-1}$ 
the wave function has a non-zero component only in layer 2 and this remains true for 
all values of the bias voltage. Due to the fact that the periodic potential is present 
only in the first layer, the wave functions and therefore also the FQHE gaps do not depend 
on the bias voltage for LL $0^{(+)}_{-1}$ and there is no phase transition in this case. 
In the LL $0^{(+)}_{-2}$, for $U=0$ the electrons are mostly located in the layer 2, 
although they have small probability of being in layer 1. Increasing the bias voltage, 
both the single-particle and the many-particle system become even more polarized in layer 2, 
and therefore the periodic potential has a negligible impact on this level as well and 
we observe the FQHE gap for all values of $U$. The situation is different for LL $0^{(+)}_{1}$ 
and $0^{(+)}_{2}$. For $U=0$ the electrons in $0^{(+)}_{1}$ are mostly localized in layer 1 
and therefore the periodic potential has a drastic impact in this case. 

In monolayer graphene the magnitude of the periodic potential $V^{}_0=20$ meV for $N^{}_e=4$ 
and $N^{}_e=6$ and for both values of $\alpha$, closeis the FQHE gap and opens the butterfly 
gap \cite{areg_butterfly}. It should be pointed out that for $\alpha=1/2$, there is no gap in 
the butterfly spectrum for the non-interacting electrons, but the electron-electron interaction 
opens a gap \cite{apalkov_14}. However, for $\alpha=1/3$ the gap is due to both the Hofstadter 
gap observed in the single-particle case and due to the contribution from the electron-electron 
interaction. Therefore, the gap for LL $0^{(+)}_{1}$ correspond to the butterfly gap for low 
values of the bias voltage $U$. The electrons in $0^{(+)}_{2}$ are mostly localized in layer 2 
and therefore the periodic potential has only a minor effect on them. Hence for the $0^{(+)}_{2}$ 
LL, the FQHE gap is observed for low values of the bias voltage $U$. By increasing the bias 
voltage $U$ there is an anticrossing between these two LLs ($0^{(+)}_{1}$ and $0^{(+)}_{2}$) 
and thereafter the layer polarization in each LL changes drastically. In particular, for the
LL $0^{(+)}_{1}$ at the bias voltage $U=250$ meV the probability of electrons being localized 
in layer 2 and for the electrons in LL $0^{(+)}_{2}$ to be localized in layer 1 are already 
$\approx0.95$. This results in a phase transition in both LLs, namely the gap in the LL $0^{(+)}_{1}$ 
which initially represented the butterfly gap now correspond to the FQHE gap. The opposite 
behavior occurs for the LL $0^{(+)}_{2}$. 

The closure of the FQHE gap by the external periodic potential also occurs in monolayer 
graphene \cite{areg_butterfly}. However, in bilayer graphene one can control the actual 
strength of the periodic potential experienced by the electrons essentially by applying 
the bias voltage. The implications of this interesting result will be discussed below. 
The described behavior is almost the same for both values of $\alpha=1/2$ and $\alpha=1/3$. 
The essential difference between these two cases is that the butterfly gap for $\alpha=1/3$ 
is substantially bigger than that for $\alpha=1/2$, and therefore the value of the bias voltage 
$U$, which is needed to observe the phase transition from the butterfly region to the FQHE region 
for the LL $0^{(+)}_{1}$ is bigger than that for $\alpha=1/2$. We have done similar studies 
for the $K'$ valley and similar phase transitions were observed in that case as well for the
LLs $0^{(-)}_{-2}$ and $0^{(-)}_{-1}$ in the negative energy region. 

\begin{figure}
\includegraphics[width=8cm]{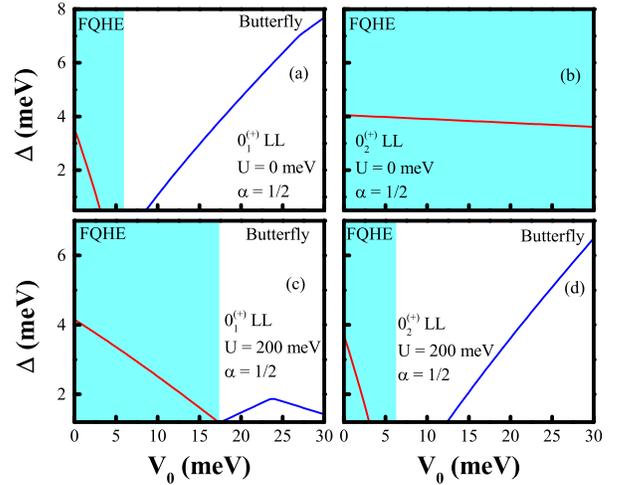}
\caption{\label{fig:GapdepV02} The dependence of both the FQHE and the butterfly gaps for 
$N^{}_e=4$ electrons on the periodic potential strength $V^{}_0$ for Landau levels 
$0^{(+)}_{1}$ and $0^{(+)}_{2}$ and for the bias voltages $U=0$ meV and $U=200$ meV, and 
$\alpha=1/2$. The line corresponding to the FQHE (butterfly) is depicted in red (blue) 
color. The region of the periodic potential strength $V^{}_0$ where the gap corresponds to 
FQHE gap is marked in cyan.}\end{figure} 

In Fig.~\ref{fig:GapdepU}, the dependence of both the FQHE and the butterfly gaps for 
$N^{}_e=4$ electrons on the bias voltage $U$ for Landau levels $0^{(+)}_{1}$ and $0^{(+)}_{2}$ 
and for two values of $\alpha$ is shown for $V^{}_0=20$ meV. The region of the bias voltage 
$U$ where the gap corresponds to the FQHE and the butterfly gap is also indicated. As was 
already pointed out, the layer polarization of the electrons changes drastically near the anticrossing 
point of the LLs $0^{(+)}_{1}$ and $0^{(+)}_{2}$ and the consequence of that can be clearly 
seen in the dependence of the gaps on the bias voltage $U$. In Fig.~\ref{fig:GapdepU}, in the 
regions further away from the anticrossing point the gaps are almost constant and fall rapidly 
to zero when approaching the anticrossing point. Also it can be clearly seen that the FQHE gap 
is almost the same for both LLs and for both values of $\alpha$, whereas the butterfly gap is 
almost twice as big for $\alpha=1/3$ compared to that of $\alpha=1/2$ as explained above.

\begin{figure}
\includegraphics[width=8cm]{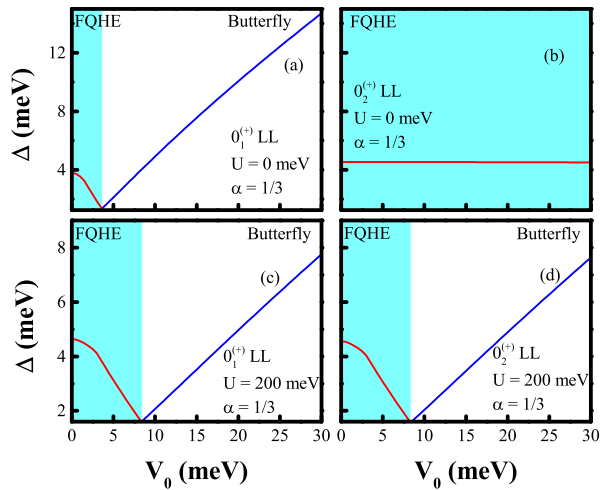}
\caption{\label{fig:GapdepV03} Same as in Fig.~\ref{fig:GapdepV02} but for $\alpha=1/3$.}
\end{figure} 

The dependence of both the FQHE and the butterfly gaps for $N^{}_e=4$ electrons on the periodic 
potential strength $V^{}_0$ for Landau levels $0^{(+)}_{1}$ and $0^{(+)}_{2}$ is shown in 
Fig.~\ref{fig:GapdepV02} for the bias voltages $U=0$ meV and $U=200$ meV, and 
for $\alpha=1/2$. The same dependence for $\alpha=1/3$ is shown in Fig.~\ref{fig:GapdepV03}. 
The region of the periodic potential strength $V^{}_0$ where the gap corresponds to the FQHE 
and the butterfly gaps is also indicated. In Fig.~\ref{fig:GapdepV02} and ~\ref{fig:GapdepV03},
similar phase transitions between the FQHE and the butterfly gaps is observed as well, 
although the dependence of the gap on the periodic potential strength $V^{}_0$ is almost linear 
in comparison to the dependence on the bias voltage $U$. No phase transition is observed for 
the LL $0^{(+)}_{2}$ and $U=0$ meV, because as was noted above, in this case the electrons are 
mostly localized in layer 2 and the periodic potential has almost no impact on the physical system. 
This feature is not observed for the case of $U=200$ meV, because as we noted above, application 
of a bias voltage gradually polarizes the electrons for the LL $0^{(+)}_{2}$ from layer 2 to layer 1 
and the effect of the periodic potential is apparent already for $U=200$ meV. A similar behavior 
is observed for LL $0^{(+)}_{1}$, where the application of the bias voltage results in widening of 
the FQHE gap region for both cases of $\alpha$. While the butterfly gap is almost linear for 
$\alpha=1/3$ in Fig.~\ref{fig:GapdepV03}, which indicates that the main contribution here is due 
to the Hofstadter gap (single particle), for $\alpha=1/2$ (Fig.~\ref{fig:GapdepV02}) it 
deviates from the linear behavior in some cases. This indicates that the butterfly gap 
due to the electron-electron interaction is highly non-trivial.
                 
While the closure of the FQHE gap and opening of the butterfly gap due to the external 
periodic potential occurs in monolayer graphene \cite{areg_butterfly}, one cannot control the periodic 
potential strength in that system. Therefore, there is no direct method to vary the periodic potential 
strength in the experiment. In bilayer graphene, as our present work indicates, this can be achieved 
by applying the bias voltage on the sample. Based on the observations above, variation of the bias voltage 
offers us the ability to control the polarization of the electrons between the two layers, which 
essentially translates to the control of the strength of the periodic potential, and at the same time 
exhibits the phase transition between the FQHE gap and the butterfly gap. This can have significant 
implications for experimental realization of the fractal butterflies in the FQHE regime.    

In conclusion, we have utilized the exact diagonalization scheme to study the FQHE and the butterfly 
gaps in bilayer graphene under the applied interlayer bias voltage and for the filling factor 
$\nu=1/3$. We have considered the case when the external periodic potential is present 
in one layer and have illustrated the effect of varying both the periodic potential 
strength and the bias voltage on the FQHE and the butterfly gaps. Two values of the parameter 
$\alpha$ were considered, namely $\alpha=1/2$ and $\alpha=1/3$. We found that by 
varying either the periodic potential strength or the bias voltage for some Landau levels in 
both valleys, a phase transition from the FQHE gap to the butterfly gap or vice versa can be 
observed. While the periodic potential strength is a characteristic of the sample used 
in the experiment and cannot be varied directly, our finding shows that by varying the
bias voltage, change of the periodic potential strength actually experienced by the 
electrons can be achieved, which can have a huge impact on the experimental investigation of 
the fractal butterflies in the FQHE region.     

The work has been supported by the Canada Research Chairs Program of the 
Government of Canada.

\end{document}